# Optical modulation of coherent phonon emission in optomechanical cavities

Jeremie Maire[1,*], Nestor E. Capuj[2,3], Martin F. Colombano[1,4], Amadeu Griol[5], Alejandro Martinez[5], Clivia M. Sotomayor-Torres[1,6], Daniel Navarro-Urrios[1,*]

[1] Catalan Institute of Nanoscience and Nanotechnology (ICN2), CSIC and BIST, Campus UAB, Bellaterra, 08193 Barcelona, Spain.
[2] Depto. Física, Universidad de La Laguna, 38200 San Cristóbal de La Laguna, Spain.
[3] Instituto Universitario de Materiales y Nanotecnología, Universidad de La Laguna, 38071 Santa Cruz de Tenerife, Spain.
[4] Depto. Física, Universidad Autonoma de Barcelona, Bellaterra, 08193 Barcelona, Spain.
[5] Nanophotonics Technology Center, Universitat Politècnica de València, 46022 València, Spain.
[6] Catalan Institute for Research and Advances Studies ICREA, 08010 Barcelona, Spain.

## Abstract

Optomechanical structures are well suited to study photon-phonon interactions, and they also turn out to be potential building blocks for phononic circuits and quantum computing. In phononic circuits, in which information is carried and processed by phonons, optomechanical structures could be used as interfaces to photons and electrons thanks to their excellent coupling efficiency. Among the components required for phononic circuits, such structures could be used to create coherent phonon sources and detectors. Complex functions other than emission or detection remain challenging and addressing a single structure in a full network proves a formidable challenge. Here, we propose and demonstrate a way to modulate the coherent emission from optomechanical crystals by external optical pumping, effectively creating a phonon switch working at ambient conditions of pressure and temperature and the working speed of which (5 MHz) is only limited by the mechanical motion of the optomechanical structure. We additionally demonstrate two other switching schemes: harmonic switching in which the mechanical mode remains active but different harmonics of the optical force are used, and switching to- and from the chaotic regime. Furthermore, the method presented here allows to select any single structure without affecting its surroundings, which is an important step towards freely controllable networks of optomechanical phonon emitters.

## Introduction

Numerous technologies have been tapping in the vast potential arising from manipulating excitations, among which electrons and photons are two of the most widespread examples. However, to use phonons in a similar way remains challenging since their coherent creation, manipulation and detection, being key elements for their practical use, prove difficult to realize. Due to the lack of discrete phonon transitions in solids at ambient conditions, major efforts have been deployed to generate coherent phonons. These phonon sources can be used for on-chip metrology and time-keeping[1] or mass/force sensing[2], amongst other applications[3]. Optomechanical (OM) crystals have become one of the most efficient way to coherently emit discrete phonons. While retarded radiation pressure[4] is the most common mechanism used in OM cavities to generate coherent vibrations at GHz frequencies[1,5–7]. Stimulated Brillouin Scattering in either waveguides or cavities has also been gaining increasing attention [8–11]. However, these two approaches impose stringent requirements on the quality of the OM structures and can thus be inconvenient to implement for room temperature operation. Recently, a new scheme for coherent phonon creation was introduced to overcome these issues. The principle to achieve coherent mechanical emission is based on a thermal/free carrier self-pulsing (SP) oscillation locking onto a mechanical mode of the structure, thus creating a self-stabilized coherent phonon source with relaxed requirements for both the optical and mechanical modes[12]. This scheme also gives access to different mechanical modes that can be brought to the lasing regime, which extend its applicability for networks of synchronised oscillators. . Furthermore, as phonons can efficiently couple to other information carrier such as photons and electrons, phononic circuits could be used as interfaces in information processing devices. In this context, sources alone are only one step towards this purpose as additional functions are required, while the most commonly sought-after feature is the synchronization of different emitters. Previous works have tackled this issue via an optical interlink [13–18], with up to three OM resonators synchronized in a cascading configuration[18] and seven[19] in the strong coupling regime. However, the potential of synchronization needed e.g., for time-keeping applications, requires the OM crystal to be in in the weak coupling regime. The aforementioned works are a first step towards synchronized networks. Nevertheless, despite these advances in synchronization, the functionalities of the emitters themselves remain limited to single mode continuous emission.

Here, we propose a scheme to address a single cavity and modulate its coherent phonon emission via external optical pumping. The mechanism behind switching is based on the shift of the resonance due to the added energy from the optical pump, which enabled us to realize a phonon switch, i.e., an on/off modulation of phonon lasing, at MHz frequencies. The coherent emission itself is achieved through the SP limit cycle described in our previous work[12,20], whereas an external optical source is used for modulation. Since the OM cavity displays other interesting regimes, such as chaos[21] which is attracting interest in various other systems[22,23], we also demonstrate state switching between these regimes, with speeds limited by the mechanical frequency. Finally, we suggest a way to extend this concept towards multi-states switches to be eventually integrated in networks of coupled oscillators.

**Materials and Methods**

The OM crystal used in this study (Fig. 1a) is fabricated from a silicon-on-insulator wafer via standard silicon (Si) nanofabrication techniques. The detailed fabrication process and geometry have already been described[12,24]. The OM structures are characterized by evanescent light coupling into the cavity from a tapered fibre at room-temperature and atmospheric pressure, as detailed in Ref.[21]. The input laser used is an infrared laser with tuneable wavelength in the range of 1440 to 1640 nm and power up to 20 mW. An 808 nm continuous-wave solid-state laser is used as an external optical pump to switch between the different states of the OM crystal. The beam passes through a pinhole to reduce its diameter, and is then modulated by a mechanical chopper at 10 kHz. The beam is then focused on the OM crystal by a 100x microscope objective, as shown in Fig. 1b and c. The temporal signals are acquired by an oscilloscope with 4 GHz bandwidth and a small mirror is used to split up part of the laser beam to trigger the acquisition.

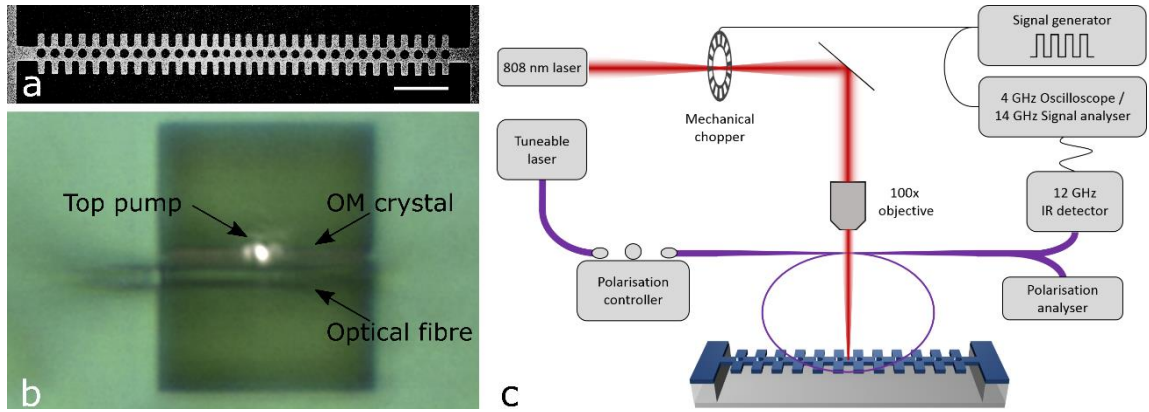

**Figure 1: Experimental details**. (a) SEM image of an OM crystal, here in shape of a suspended Si nanobeam. Scale bar is 2 µm. (b) Optical image showing a top view of the OM crystal with the fibre below and the pump laser spot. The length of the nanobeam is 16.5 µm. (c) Schematic of the evanescent fibre coupling measurement system.

We first characterize the basic optical and mechanical properties of the OM structure. For that purpose, we start by acquiring a low-power (0.2 mW) optical transmission spectrum which displays a number of resonances. As the OM coupling coefficient is higher for low-order modes, we usually focus on the first resonance to excite mechanical vibrations. Increasing the optical input power results in a "saw-tooth" shaped transmission, as the resonant frequency is red-shifted due to the contribution of the thermo-optic (TO) effect[25]. The mechanical spectrum is obtained by processing the transmitted optical signal with a spectrum analyser. The initial state corresponds to a large laser-cavity blue-detuning, i.e., the wavelength of the laser is shorter than the wavelength of the resonance, and then the input wavelength is progressively increased. For a large detuning, the string-like mechanical modes are thermally driven and appear as Lorentzian peaks in the transmission spectra, represented by blue lines on the left of the dashed line in the map of Figure 2, with quality factors on the order of $10^2$. Due to symmetry considerations[12], we focus on the second order (three anti-nodes) mechanical odd mode of frequency $v_{m,2}$ = 54 MHz. Decreasing the detuning brings the OM crystal in a SP limit cycle[12,26–30] that stems from the interplay between the free-carrier dispersion (FCD) and the TO effect. As the input wavelength is further increased, the frequency of the SP limit cycle increases until one of the higher harmonics of the optical force reaches the frequency of the mechanical mode. At this point the SP frequency locks onto that of the mechanical mode, leading to its amplification and resulting in coherent, high-amplitude mechanical vibrations, qualified as phonon lasing[12]. Increasing further the wavelength can lead to coherent activation of other mechanical states. For specific values of the input laser parameters, a chaotic regime can be reached[21]. Examples of RF spectra maps are displayed in Figure 2c and d. Below, we explain the principle of the switching mechanism, followed by our experimental observations and describe the characteristics of the phonon switch realised via the external optical pumping scheme.

## Results and Discussion

### Switching mechanism

To explain the mechanism of switching, regardless of the regime involved, let us first look at the "pump off" state. The optical resonance that we use displays an asymmetric shape due to the thermo-optic effect, as shown in Fig. 2a. During the experiment, the input laser is fixed at a specific wavelength, which corresponds to one of the regimes of the structure, e.g. self-pulsing, phonon lasing or chaos in Fig. 2a. When activating the external optical switching, the external pump acts as a source of energy for the OM crystal, hence increasing its temperature and enhancing the thermo-optic effect. This will in turn shift the resonance to higher wavelengths, as shown in Fig. 2b. Comparatively, the input wavelength in the fibre, which is already fixed, will be further away from the maximum wavelength of the resonance. In the example given in Fig. 2a and b, this brings the system to the SP regime (pump on) from the initial mechanical lasing regime (pump off). To confirm this phenomenon, we acquire two maps of RF spectra as a function of wavelength, one corresponding to the initial condition, i.e. without additional pumping (Fig. 2c) and a second with external pumping (Fig. 2d). It is clear that the resonance displays similar regimes, including mechanical lasing, but shifted to higher wavelengths when additional pumping is introduced, in agreement with the previous explanation. Hence, in the switching experiments presented hereafter, the initial state (pump off) always corresponds to the regime requiring the highest input wavelength. When turning on the pump, the system then reverts to a state requiring, in principle, a lower wavelength. Note that the input wavelength in the optical fibre is fixed during each experiment.

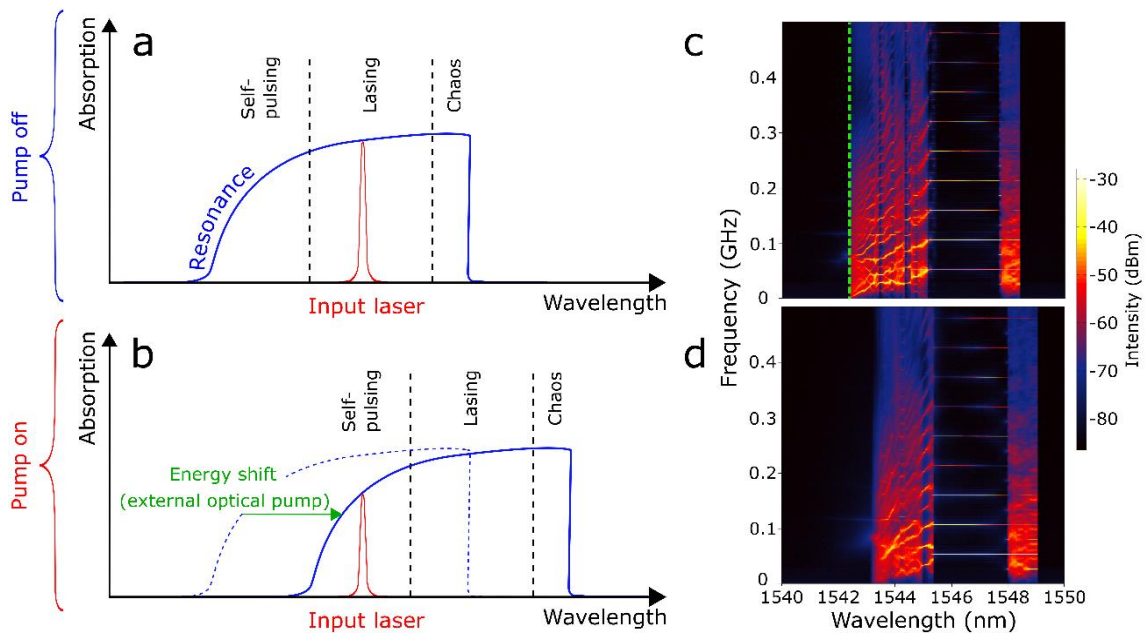

**Figure 2: Switching mechanism**. (a) Diagram of the switching experiment. The resonance is shown as a function of wavelength. The asymmetry stems from the thermo-optic effect and the three regions within the resonance are given as an example. The input laser is used to bring the system to the mechanical lasing regime. (b)The external optical pump is switched on, shifting the resonance to higher wavelengths. The input laser remains at the same wavelength as in the "pump off" state, hence the change in the working regime of the emitter. (c) Map of RF spectra as a function of wavelength, around a specific optical resonance, without external pumping. The dashed line marks the limit between the thermally driven mechanical modes (left) and the SP regime. (d) Identical resonance recorded with external pumping switched on. A clear shift towards longer wavelengths is observed as compared to (c), as per the diagram in (a).

**Phonon switch**

In all the subsequent switching experiments, the excitation laser is used to bring the system to the initial state, after which wavelength and power remain fixed, whereas the optical pump is used to trigger the switching. We first drive the system, without external pumping, to the mechanical lasing regime at 54 MHz, the spectrum of which is displayed in Fig. 3a (top). The optical pump is then turned on, effectively switching off the mechanical lasing, as shown by the corresponding spectrum in Fig. 3a (bottom). To identify the two states with more precision, the temporal signals shown in Fig. 3b are examined. Whereas the on-state of the external pumping leads to a very weak modulation of the signal, the initial state shows a comb of harmonics of the main RF peak typical of the phonon lasing regime driven by SP, fully described in our previous work[12]. Note that although Fig. 3a (bottom) shows the 54 MHz mechanical mode with low amplitude, we have also observed the switching between the lasing regime and a pump-on state in which the spectrum does not display anything except for the background noise (see Supplementary video 1). With the on/off states clearly identified, we then focus on the switching mechanism itself. For that purpose, we record high-resolution temporal signals of both the external optical pump and the signal from the OM structure, both shown in Fig. 3c. The normalized transmission clearly shows a periodic signal with two distinct regions: a low amplitude signal when the external pump is on and high amplitude oscillations when the pump is off.

Such a switch has two main characteristics: the switching speed and its switching power. To characterize the switching speed, we take a close-up look at the on/off transitions in Fig. 3d-e, recorded at a repetition rate of 10 kHz. After a transition is initiated, we observe that approximately 200 ns are needed for the signal to stabilize in its new state, giving a switching rate of 5 MHz, which has to be compared to the mechanical frequency of 54 MHz. There are mainly two limiting factors to this switching speed. First, the speed of the On/Off transition of the optical pump signal is limited by the use of a mechanical chopper. Second, the mechanical frequency itself limits the switching speed as we expect a few mechanical periods to be necessary to establish amplified mechanical vibrations. Hence the switching speed of 5 MHz is a lower bound to the actual value.

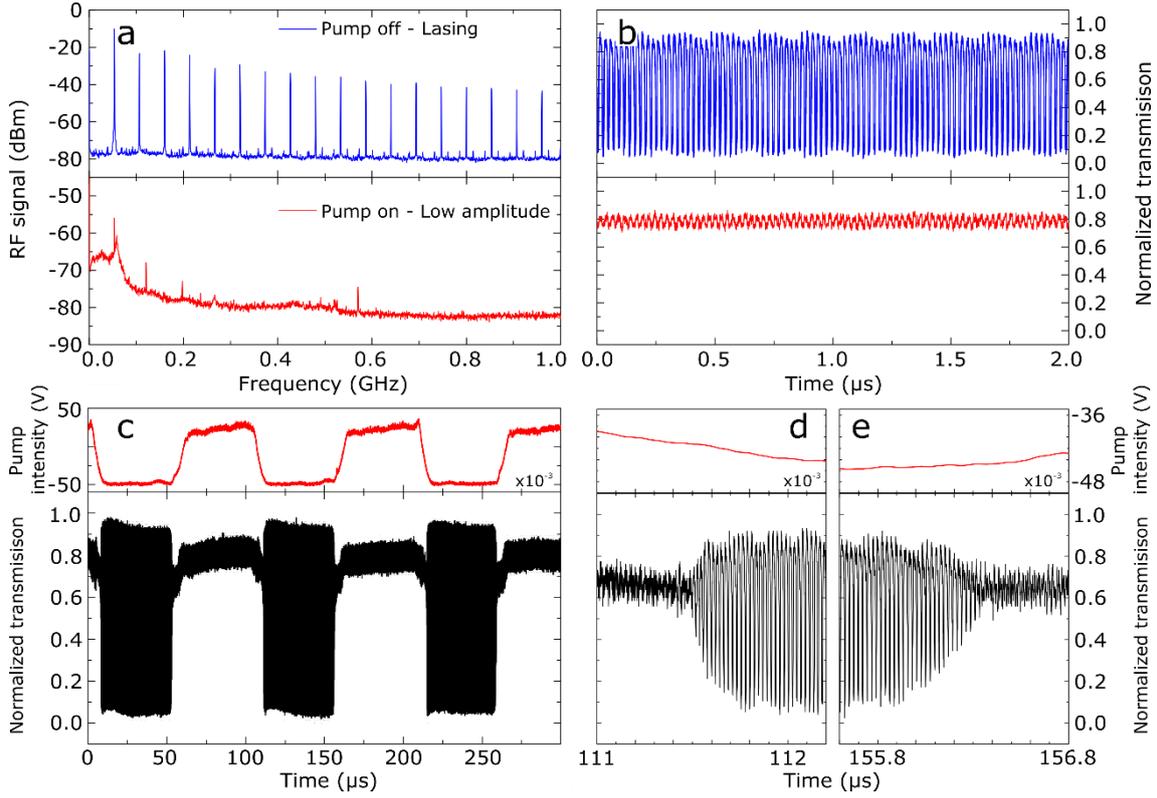

**Figure 3: Phonon switch**. (a) RF spectra in the initial lasing regime (pump off, top) and in a low mechanical amplitude regime with self-pulsing (pump on, bottom). (b) Corresponding temporal signal for the lasing (top) and SP (bottom) regimes. (c) Large scale temporal signal displaying the external pumping signal (top) and the corresponding signal from the OM crystal. (d-e) Zoom on the temporal signals of the on→off and off→on transitions of the phonon switch, respectively.

To address the switching power, we monotonically increase the input laser wavelength until phonon lasing is observed. We verify, as in Fig. 3, that the switch is stable by cycling the pump on and off. In this situation, the lowest pump power at which we observed switching was 130 µW. This power is measured just after the microscope objective and represent the power needed by our experiment to induce switching, which is higher than the power needed inside the OM crystal. Considering that the power input through the fibre is typically between 5 and 10 mW, the switching power represents less than 3% of the input laser power, which is a very promising power efficiency.

**Switching between other regimes**

A number of regimes are available in the OM crystal under study. Therefore, we investigate the additional switching possibilities available. In the switching scheme presented before, the first harmonic of the optical force (in the SP regime) is used to bring the system to mechanical lasing. In a previous work, we showed that it is possible to use higher harmonics to bring the same mechanical mode to lasing as well[12]. Hence, we perform switching experiments between the 1st (M=1) and 2nd (M=2) harmonic driving of the mechanical mode (Fig. 4). We observe that this transition occurs instantaneously, i.e., there is no discontinuity in the mechanical lasing temporal trace, as shown in inset of Figure 4. In this case, the mechanical mode is already active. Hence, the transition only involves the SP phenomena, i.e., thermo-optic and free carrier dispersion. Since the change occurs much faster than the relatively long period of the mechanical mode, the transition appears immediate.

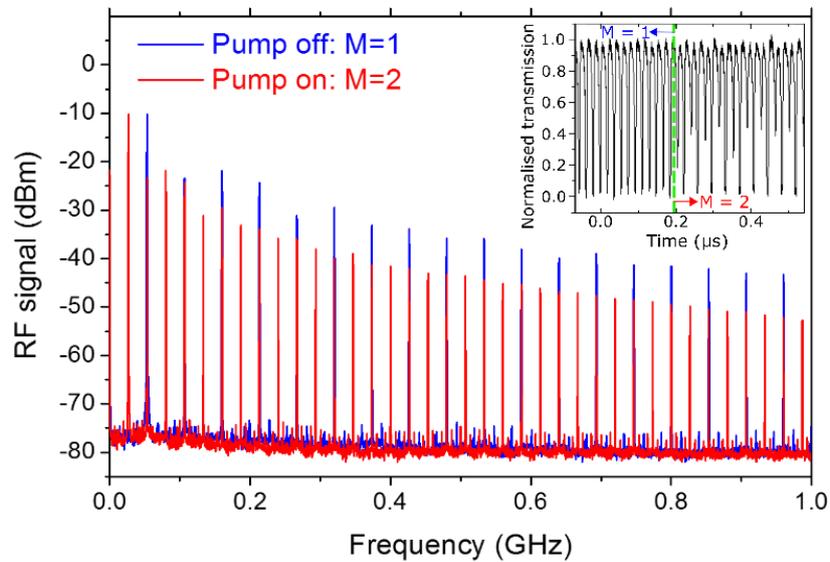

**Figure 4: Harmonic switching**. RF spectra in the phonon lasing regime driven by the first (M=1) and second (M=2) harmonics of the optical force. (Inset) Temporal signal of the transition between the pump off state (M=1) and the pump on state (M=2).

More interestingly, we focus on switching involving a chaotic state, as this regime could have applications in secure data communications or sensing. Fig. 5a and b show the spectra and temporal signals in the chaotic (top) and phonon lasing (bottom) regimes. Note that numerical simulations suggest that the chaotic behaviour does not stem from chaotic mechanical vibrations[21]. To investigate the switching mechanism, the system is initially put in the chaotic state and the external optical pumping is then applied. In a similar way to the phonon switch, under external pumping the system reverts back to a state requiring shorter input wavelength, in this case phonon lasing. Due to the complex nonlinear signals, it is difficult to precisely characterize the switching speed, but from inspection of the transitions in Fig. 5c, the exclusively optical nature of the switching involved, given that the mechanical mode stays active in both the on and off states, enables extremely fast switching, i.e., fast enough for the transition to occur within one period of the mechanical vibration.

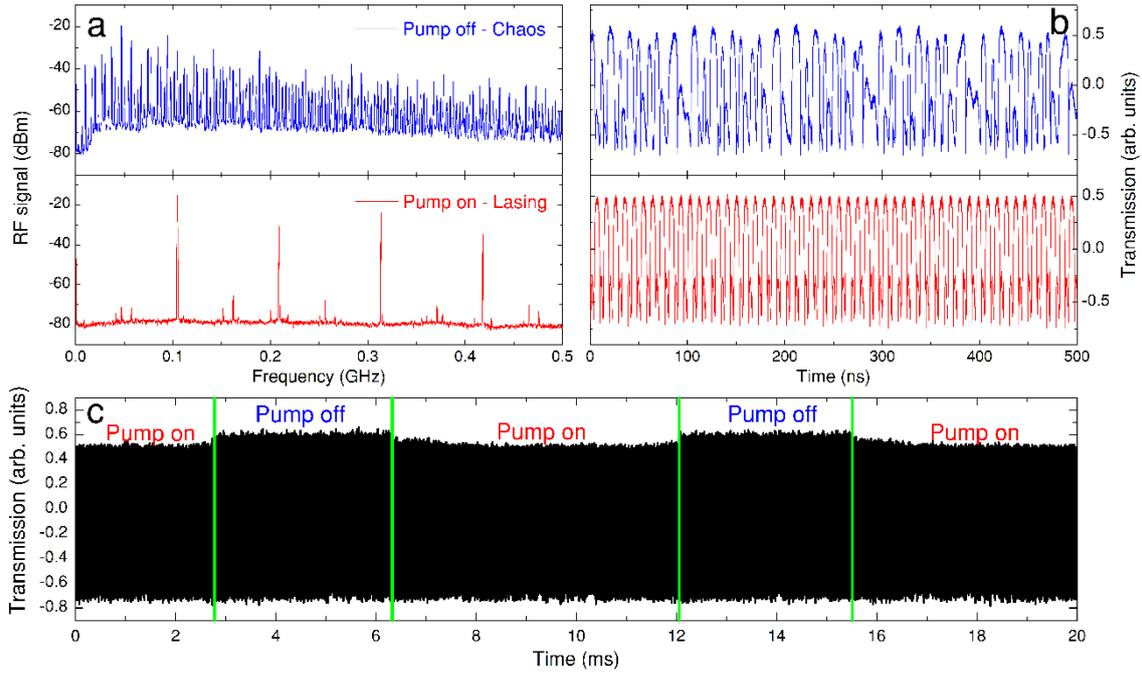

**Figure 5: Chaos-Lasing switching.** (a) RF spectra in the initial chaotic regime (pump off, top) and in the phonon lasing regime (pump on, bottom). (b) Corresponding temporal signals for the chaotic (top) and lasing (bottom) regimes. (c) Large scale temporal signal displaying both on and off regions of the optical pumping.

## Conclusions

The three switching mechanisms presented in this work constitute proof of concepts of phonon switching, one of the crucial functions to implement phononic circuitry. We have shown the possibility to address a single OM cavity and modulate its coherent phonon emission at a speed above 5 MHz. This feature will become increasingly important as the number of phonon emitters per unit area increases. In addition, this method is tailored specifically for the phonon lasing scheme of this work and is not affected by ambient temperature and pressure conditions. The phonon switching is energy efficient since the power required is less than 3% of that used to reach the phonon lasing regime. In order to increase the possibilities of the switching function, different paths are available. One is to engineer the resonance, so that it displays a larger number of stable states for different input wavelengths. It is also possible to take advantage of phenomena such as multistability to go beyond a binary switch. Indeed, bistability in this type of OM cavities has already been demonstrated for both power and wavelength[21]. Finally, a very promising route is to intercouple two or more similar OM cavities either with mechanical or optical weak links and investigate the switching functions demonstrated in this work in the context of synchronized oscillators. Therefore, this work is a promising step towards the realisation of phononic circuit functions in OM crystals.

**Acknowledgements**

This work was supported by the European Comission FET Open project PHENOMEN (H2020-EU-713450) and the Spanish MINECO project PHENTOM (FIS2015-70862-P). The ICN2 is funded by the CERCA programme, the Generalitat de Catalunya and supported by the Severo Ochoa programme of the Spanish MINECO grant no. SEV-2013-0295. DNU and MFC gratefully acknowledge the support of a Ramón y Cajal postdoctoral fellowship (RYC-2014-15392) and a Severo Ochoa studentship, respectively. The authors thank M. Sledzinska, P. D. Garcia, G. Arregui and F. Alzina for fruitful discussions. **Author contributions:** J.M., M.F.C and D.N.-U. performed the experiments. J.M., D.N.-U. and N.E.C. analysed the data. A.G. and A.M. fabricated the samples. D.N.-U., N.E.C and C.M.S.-T. conceived the idea of this work. All authors contributed to the interpretation of the results and writing of the manuscript.